\documentclass[12pt]{article}
\usepackage[latin1]{inputenc}
\usepackage{amsmath}
\usepackage{amssymb}
\usepackage{amsfonts}
\usepackage{amssymb}
\usepackage{graphicx}
\usepackage{footnote}
\usepackage{float}
\usepackage{subfigure}
\DeclareGraphicsExtensions{.bmp,.png,.pdf,.jpg,.eps}
\usepackage{physics}

\usepackage[colorlinks]{hyperref}
\hypersetup{
    colorlinks=true,
    linkcolor=blue,
    filecolor=magenta,  
    urlcolor=[rgb]{0.00,0.00,0.50},    
    citecolor=red,
    }

\usepackage{url}

\usepackage{amsmath}
\usepackage[english]{babel}
\usepackage[latin1]{inputenc}
\usepackage{amsmath}
\numberwithin{equation}{section}

\usepackage{amsmath}
\usepackage[english]{babel}

\def\be{\begin{equation}}
\def\ee{\end{equation}}

\def\Z{\mathbb Z}

\def\R{\mathbb R}

\usepackage{todonotes}

\begin{document}

\title{{\Huge {\bf Exotic nonlinear supersymmetry and  integrable systems}}}

\author{{\bf  Mikhail S. Plyushchay} 
 \\
[8pt]
{\small \textit{Departamento de F\'{\i}sica,
Universidad de Santiago de Chile, Casilla 307, Santiago,
Chile}}\\
[4pt]
 \sl{\small{E-mail:
\textcolor{blue}{mikhail.plyushchay@usach.cl}
}}
}
\date{}
\maketitle

\begin{abstract}
Peculiar properties of many classical and quantum systems 
can be related to, or derived from those of a free particle.
In this way we explain the appearance and peculiarities of the exotic nonlinear  
Poincar\'e supersymmetry  in reflectionless and  finite-gap   quantum systems
related to the Korteweg-de Vries equation.
The same approach is used to  explain the origin 
and the nature of nonlinear
 symmetries in the perfectly invisible  $\mathcal{PT}$-regularized conformal 
 and superconformal  mechanics systems.
\end{abstract}

Peculiar properties of many classical and quantum systems can be related to, 
or derived  from those of a  free particle.
The peculiarity  of the simplest  
1D free particle system is that it has 
a local integral of motion $p=-i\frac{d}{dx}$, which separates 
the left- and right-moving plane waves $e^{\pm ikx}$ of the same energy, 
and distinguishes  the unique 
non-degenerate   state $\psi_0=1$ of zero energy 
at the very edge of the continuous spectrum by   annihilating it. 
This simplest system also is characterized  by the $\mathfrak{sl}(2,\R)$ conformal symmetry that 
expands up to the Schr\"odinger symmetry due to the presence
of the integral $p$ \cite{Nied}.


The interesting class of the quantum systems intimately related to 
the simplest case of the one-dimensional free particle corresponds to 
reflectionless systems \cite{KayMos}. 
Applying a generalized Darboux transformation \cite{MatSal}
of order $n\geq 1$ 
to the quantum 1D free particle $H_0$,
a quantum reflectionless  system $H_n$ is generated. 
Each  energy level in the continuous part of the spectrum of $H_n$
corresponds to a deformed plane wave $\Psi_{\pm k}(x)=\mathcal{A}^-e^{\pm ikx}$
 propagating to
the left or to the right. The generator 
of the  Darboux transformation 
 $\mathcal{A}^-$ is a differential operator 
of order $n$ 
based here on  an appropriately  chosen set of the seed states 
$\psi^s_j(x)$, $j=1,\ldots, n$,  which are
 formal (non-physical) eigenstates of $H_0$,
 $\text{ker}\,\mathcal{A}^-=\text{span}\,\{\psi^s_1,\ldots,\psi^s_n\}$.
 Reflectionless system  $H_n$ has  then $n$
bound states $\Psi_{j}(x)=\mathcal{A}^-\widetilde{\psi^s_j}$,
$j=1,\dots, n$, generated from 
linearly independent formal eigenstates $\widetilde{\psi^s_j}(x)=\psi^s_j(x)\int^xd\xi/(
\psi^s_j(\xi))^2$ of $H_0$ of the same eigenvalues as the states $\psi^s_j$
 \cite{InzPlyNu}.
 A non-degenerate state $\Psi_0(x)=\mathcal{A}^-1$ 
at the very edge of the continuos part of the
 spectrum is a Darboux-transformed non-degenerate state 
$\psi_0=1$ of the free particle.  All these spectral peculiarities of reflectionless system $H_n$ 
are detected by a nontrivial integral of motion $\mathcal{P}=\mathcal{A}^-p\mathcal{A}^+$,
$\mathcal{A}^+=(\mathcal{A}^-)^\dagger$, 
 which is a differential operator of the odd order $2n+1$ being 
 a Darboux-dressed momentum operator of the free particle.
 The operator $\mathcal{P}$ annihilates all the $n$ bound states 
 as well as the lowest state $\Psi_0(x)$ in the continuos part of the spectrum of $H_n$,
 and separates the  left- and right-moving deformed plane waves $\Psi_{\pm k}(x)$ of 
 equal energy 
 being eigenstates of the $\mathcal{P}$ of opposite eigenvalues 
 \cite{ArMatPly1}. 
 
 
Potentials of the quantum reflectionless systems can be promoted to multi-soliton solutions 
of the Korteweg-de Vries (KdV) equation by exploiting  the covariance of its Lax representation
with respect to  the Darboux transformations 
\cite{MatSal,Defects2}. 
In this way,  potential of any reflectionless system with $n$ bound states 
represents a snapshot of an $n$-soliton solution to the KdV equation, whose temporal evolution
corresponds  to an isospectral deformation of the reflectionless system.
Operator $\mathcal{P}$ in such a picture is a Lax-Novikov integral of the $n$-th  stationary equation 
of the KdV hierarchy 
\cite{ArMatPly1}.

       
   By periodization of reflectionless systems,   some  finite-gap 
quantum systems can be obtained, whose potentials   are solutions of the stationary 
equations of the KdV hierarchy. 
 Lax-Novikov integral of an $n$-gap  quantum system, being differential operator of order $2n+1$, 
separates the left- and right-moving Bloch states of the same energy  inside 
the valence and conduction bands, and  annihilates all the $2n+1$ periodic and anti-periodic 
edge states 
at the edges of the bands, on which two irreducible non-unitary finite-dimensional
representations of the conformal $\mathfrak{sl}(2,\R)$ algebra are realized 
\cite{CJNP}.
The Darboux covariance of the Lax representation allows to promote the potentials 
of finite-gap quantum systems  to the  cnoidal-type solutions of the KdV equation 
\cite{Defects2}. 


      Darboux transformations also can be applied to the finite-gap systems  
      of the most general form to produce some finite-gap systems 
      completely isospectral to the initial ones, or to generate finite-gap systems with 
      the added  arbitrary number of the bound states 
      inside the  prohibited zones or at their edges. In the latter case the generated  potentials 
      and related to them super-potentials 
      are promoted to  solutions of  the   KdV and the modified KdV  equations in the form 
      of the soliton defects propagating in a finite-gap background  
      \cite{Defects2}.

 
   With the pairs of the  quantum systems produced  starting 
   from the  quantum 1D free particle or a finite-gap system, 
   an  exotic  nonlinear $\mathcal{N}=4$ supersymmetry can be associated.
The emergence of the exotic supersymmetry in  such systems 
is rooted in existence of the momentum integral  in the free particle,
or the Lax-Novikov  integral in a  finite-gap system.
In the simplest case of a reflectionless system, due to the presence of $p$
in the structure of  the  Lax-Novikov integral $\mathcal{P}$, the latter  can be 
factorized  into  the product of two non-singular operators,
 $\mathcal{P}=\mathcal{A}^-(p\mathcal{A}^+)$.
In correspondence with this, the given reflectionless system $H_n$
can be generated from and intertwined with
the free particle $H_0$ not only by the 
order $n$ differential 
operators $\mathcal{A}^-$ and $\mathcal{A}^+$, but also 
by the operators  $\mathcal{A}^-p$ and $p\mathcal{A}^+$
of differential order $n+1$.  
As a consequence, the extended system $\mathcal{H}$ 
composed from $H_0$ and $H_n$ has not only a pair of 
supercharges $Q_a$, $a=1,2$,
constructed from the operators $\mathcal{A}^-$ and $\mathcal{A}^+$,
but also possesses a pair of supercharges $S_a$ of differential order $n+1$
 constructed from the intertwining operators
 $\mathcal{A}^-p$ and $p\mathcal{A}^+$. The anti-commutators 
 of  the supercharges  $Q_a$ and $Q_b$ produce a polynomial 
 of order $n$ in $\mathcal{H}$, while  
 $S_a$ and $S_b$ anti-commute for a polynomial 
 of order $n+1$ in $\mathcal{H}$. The anti-commutator of 
 $Q_a$ and $S_b$ gives rise to  an additional even generator $\mathcal{L}$ of
 the superalgebra composed from $pH_0^n$ and Lax-Novikov integral
 $\mathcal{P}$. As a result, 
 instead of  the $\mathcal{N}=2$ (nonlinear in the case of $n>1$)  
Poincar\'e supersymmetry generated by two supercharges and Hamiltonian
$\mathcal{H}$,
we obtain an exotic nonlinear $\mathcal{N}=4$
Poincar\'e supersymmetry which includes an additional 
bosonic integral $\mathcal{L}$.
Analogous exotic nonlinear $\mathcal{N}=4$
Poincar\'e supersymmetric structure describes extended systems $\mathcal{H}$ 
composed from  isospectral, or almost isospectral pairs 
of reflectionless systems with 
multi-soliton potentials $u_n(x,\ldots)$ and $ u_{n'}(x,\ldots)$,
where the ellipsis corresponds to the sets of $2n$ and $2n'$ 
parameters characterizing the amplitudes and phases of 
the $n$- and $n'$-soliton solutions of the KdV equation.
The concrete form of the superalgebra depends on the choice 
of those parameters,  and its 
 supercharges undergo some restructuring
  associated with  lowering  their differential orders each time when some sets of the 
 amplitude parameters in $H_n$ coincide with those in super-partner  reflectionless system 
 $H_{n'}$ 
 \cite{ArMatPly1}. 
 Additional restructuring in 
 supercharges
 and  exotic nonlinear superalgebra generated by them  can
 also happen 
 for special values of the phase differences  associated 
 with the coinciding pairs of the soliton amplitudes in potentials 
 $u_n(x,\ldots)$ and $ u_{n'}(x,\ldots)$
  \cite{ArMatPly1}.
 In all the cases, however, 
 a pair of supercharges are matrix  operators of 
 some even differential order, while another pair of supercharges
 has an odd differential order. This is related to the nature of the Lax-Novikov 
 integrals of the subsystems $H_n$ and $H_{n'}$, which  
 have odd differential orders $2n+1$
 and $2n'+1$, and that supercharges from different pairs
 effectively provide the factorization of Lax-Novikov integrals
 into two non-singular differential
 operators.
 For such extended quantum systems $\mathcal{H}$, the phenomenon of transmutation
 between the exact 
 and partially broken exotic nonlinear supersymmetries 
 was observed and   interpreted   in terms of the soliton 
 scattering  in 
 \cite{AraPlyTrans}. 
 In the case of the unbroken supersymmetry, 
 the unique ground state of the system 
  $\mathcal{H}$  is a zero mode of all the odd generators 
  $Q_a$ and $S_a$ and of the even generators   $\mathcal{H}$ and 
  $\mathcal{L}$ of the superalgebra. Coherently with this,  the  operator
    $\mathcal{L}$ is a central element of the exotic nonlinear 
    $\mathcal{N}=4$ Poincar\'e superalgebra.  In the phase of the partially broken
    exotic nonlinear supersymmetry,  the even 
   generator $\mathcal{L}$ mutually transforms  the pairs 
   of the supercharges $Q_a$ and $S_a$  by means of the commutator, 
   and the system   $\mathcal{H}$
   has a doubly degenerate lowest energy level,
   whose corresponding states 
   are annihilated by a part of the  supercharges 
   \cite{ArMatPly1,AraPlyTrans}.
   

A similar exotic nonlinear $\mathcal{N}=4$
Poincar\'e supersymmetric structure also 
describes extended systems $\mathcal{H}$ 
composed from  isospectral  pairs 
of the finite-gap quantum  systems,
and finite-gap systems with soliton 
defects.  In the latter case, the fine structure of
the exotic supersymmetry  controls the  nature 
and propagation of soliton  defects with energies 
which can  be introduced 
into different prohibited zones of the finite-gap systems 
\cite{Defects2}.

    
The KdV equation has also rational  solutions, in which 
the dynamics of the moving poles 
is governed  by the Calogero-Moser systems.
Such solutions   can be obtained via an appropriate  limit 
procedure from multi-soliton solutions 
by  exploiting  the Galilean symmetry of the KdV equation.
They  also can be obtained directly from the free particle 
by applying to it  singular generalized Darboux transformations 
based on zero-energy eigenstates $\psi_0=1$ and $\widetilde{\psi_0}(x)=x$
 of the free particle and Jordan  states corresponding to the same zero 
 energy 
 \cite{MatPly}. 
The simplest case of the Darboux-generated  in this way system
corresponds to the two-particle  Calogero model with the 
omitted center of mass coordinate. Due to a singular nature of the Darboux transformation,
the  Schr\"odinger symmetry of a free particle  reduces and transforms  
into conformal 
  $\mathfrak{sl}(2,\R)$ symmetry of the generated Calogero system with a non-degenerate 
  continuos spectrum $(0,\infty)$. Though the Darboux-dressed momentum operator
  $\mathcal{P}=\mathcal{A}^-p\mathcal{A}^+$ in this case commutes with 
  the generated Hamiltonian $H_n$, it is a formal, non-physical integral of motion
  since acting on non-degenerate eigenstates of $H_n$ it transforms
  them into  non-physical, formal  eigenstates of $H_n$ which do not
  satisfy the Dirichlet boundary condition at $x=0$ 
  \cite{COP}. 
  Coherently with a non-physical nature of the formal Lax-Novikov integral,
  the corresponding extended system $\mathcal{H}$   is described 
  by (a  non-linear in general case) $\mathcal{N}=2$ 
  Poincar\'e supersymmetry generated by supercharges $Q_a$ 
  constructed from the 
  operators $\mathcal{A}^-$ and  $\mathcal{A}^+$,
 but the  exotic $\mathcal{N}=4$ 
  Poincar\'e supersymmetry is lost.
  
  
  Then the natural question arises whether it is possible to somehow 
  restore the exotic nonlinear $\mathcal{N}=4$ 
 Poincar\'e supersymmetry in the 
 quantum systems associated with 
 rational solutions of the KdV equation.
 

 In \cite{MatPly} it was  recently shown that   this indeed can be achieved via the 
 $\mathcal{PT}$-regularization 
$x\rightarrow x+  i\alpha$, $\alpha\in \R$, $\alpha\neq 0$, of Darboux transformations. 
The key point of such a complex shift is that it allows to recuperate 
 the Schr\"odinger symmetry in a Darboux-generated system,
 where, however, a higher derivative Lax-Novikov 
 integral expands its algebra  and  transforms into a non-linear one. 
 The obtained in such a way systems 
 possess several interesting properties. They are not only refectioness,
 but are perfectly invisible since in them
  the transmission amplitude itself, and not only its modulus,
   is equal to one. Another peculiarity is that each of them  contains 
   a unique bound state of zero energy at the very edge of the continuous part
   of the  spectrum, which is  described by a quadratically integrable wave function,
   and in this sense they are zero-gap quantum systems.
  The paired  perfectly invisible systems are described by 
  different forms  of the nonlinearly extended   generalized super-Schr\"odinger 
  symmetry, which can include or not include  the superconformal
  $\mathfrak{osp}(2,2)$ symmetry in the form of a sub-superalgebra
  depending on  the unbroken or partially broken phase of 
  the exotic nonlinear $\mathcal{N}=4$ Poincar\'e supersymmetry in them.
  The potentials of such perfectly invisible  $\mathcal{PT}$-invariant 
quantum systems can be promoted to the solutions 
of the complexified  KdV equation (or higher equations of the hierarchy), 
which  exhibit, particularly, 
a behaviour typical for extreme (rogue) waves.

   
The simplest system with the unbroken 
exotic $\mathcal{N}=4$ nonlinear supersymmetry is described by the Hamiltonian 
and supercharges
\be\nonumber
\mathcal{H}=
\left(
\begin{array}{cc}
 H^\alpha_1 & 0   \\
0  &  H_0
\end{array}
\right)\,,\qquad
Q_1=
\left(
\begin{array}{cc}
 0 & D_1   \\
D^\#_1  &  0
\end{array}
\right),\qquad
S_1=
\left(
\begin{array}{cc}
 0 & -iD_1 \mathcal{P}_0  \\
i\mathcal{P}_0 D^\#_1  &  0
\end{array}
\right)\,,
\ee
 and $Q_2=i\sigma_3Q_1$, 
$S_2=i\sigma_3S_1$.
Here $\mathcal{P}_0=p=-i\frac{d}{dx}$ is the
momentum operator of the free particle $H_0=-\frac{d^2}{dx^2}$,
$D_1=\xi\frac{d}{dx}\xi^{-1}=\frac{d}{dx}-\xi^{-1}$
and 
$D^\#_1=-\xi^{-1}\frac{d}{dx}\xi=-\frac{d}{dx}-\xi^{-1}$
are constructed  on the base of 
$\xi=x+i\alpha$ which is a non-physical  
zero-energy eigenstate of $H_0$.  Operators $D_1$ and $D^\#_1$
are the Darboux generators  $\mathcal{A}^-$ and $\mathcal{A}^+$ 
for the super-partners 
$H_0$
and $ H^\alpha_1=-\frac{d^2}{dx^2}+2\xi^{-2}$.
The $\mathcal{H}$, $Q_a$ and $S_a$   generate the non-linear 
superalgebra $[\mathcal{H},Q_a]=[\mathcal{H},S_a]=0$, 
\be\nonumber
\{Q_a,Q_b\}=2\delta_{ab}\mathcal{H}\,,\qquad
\{S_a,S_b\}=2\delta_{ab}\mathcal{H}^2\,,
\qquad
\{Q_a,S_b\}=2\epsilon_{ab}\mathcal{L}\,,
\ee
where
\be\nonumber
\mathcal{L}=
\left(
\begin{array}{cc}
 \mathcal{P}^\alpha_1 & 0  \\
0 &  H_0\mathcal{P}_0
\end{array}
\right)
\ee
is the bosonic integral of motion being  a central charge of this superalgebra.
The kernel of the Lax-Novikov integral $ \mathcal{P}^\alpha_1=D_1\mathcal{P}_0 D^\#_1 $ 
of the $\mathcal{PT}$-regularized two-particle Calogero  subsystem 
$ H^\alpha_1$
is
$    {\rm ker}\, \mathcal{P}_1^\alpha
    =\text{span}\, \{\xi^{-1},\xi,\xi^3\}\,$.
Here $\xi^{-1}$ is the zero-energy bound state of $ H^\alpha_1$,  while  $\xi$ and 
 $\xi^3$ are its Jordan states,  $H^\alpha_1\xi=2\xi^{-1}$, $H^\alpha_1\xi^3=-4\xi$.
 The unique ground state of the system $\mathcal{H}$ of zero-energy 
 $\Psi_0=(D_1 1,0)^t=(-\xi^{-1},0)^t$ is annihilated by
 all the supercharges $Q_a$ and $S_a$ as well as by 
 the even generator $\mathcal{L}$. 
 
 The set of the even operators $\mathcal{H}$, $(\mathcal{I}-\frac{1}{2}\Sigma)$,
 $\mathcal{K}=\text{diag}\,({K}_1^\alpha, {K}_0^\alpha)$,
$\mathcal{D}=\text{diag}\,({D}_1^\alpha, {D}_0^\alpha)$, and odd 
operators $Q_a$ and $\lambda_1=-\xi\sigma_2-2tQ_1$, 
$\lambda_2=i\sigma_3\lambda_1$
generate the $\mathfrak{osp}(2\vert 2)$ superalgebra of 
the matrix system $\mathcal{H}$. Here $\mathcal{I}=\text{diag}\,(1,1)$ and $\Sigma=\sigma_3$;
${D}_0^\alpha=\frac{1}{4}\{G^\alpha_0,\mathcal{P}_0\}$
and 
${K}_0^\alpha=(G^\alpha_0)^2$ are the generators of conformal $\mathfrak{sl}(2,\R)$
symmetry of $H_0$
being its time-dependent, dynamical integrals of motion
constructed on the base of its generator 
of Galileo transformations $G^\alpha_0=\xi-2t\mathcal{P}_0$, while
${D}_1^\alpha=\frac{1}{4}\{\xi,\mathcal{P}_0\}-tH^\alpha_1$ and 
${K}_1^\alpha=\xi^2-8t{D}_1^\alpha-4t^2H^\alpha_1$
are the analogous $\mathfrak{sl}(2,\R)$ generators for $H^\alpha_1$.
Extension of the set of the generators of  superconformal $\mathfrak{osp}(2\vert 2)$ 
symmetry of the system $\mathcal{H}$ by 
the even integral $\mathcal{L}$ gives rise to the 
expansion  of the set of the integrals of motion 
by the set
of the even integrals 
\begin{eqnarray}\nonumber
&
\Sigma\,,\qquad  \mathcal{P_-}=\frac{1}{2}(1-\sigma_3)\mathcal{P}_0\,,
\qquad
\mathcal{G_-}=\frac{1}{2}(1-\sigma_3){G}_0^\alpha\,,
\qquad
\mathcal{G}=\text{diag}\,\left(G_1^\alpha,\,  \frac{1}{2}\{G_0^\alpha,H_0\}\right)\,,
&
\\
&
\mathcal{V}=
i\xi^2D^\#_1\mathcal{I}-4t\mathcal{G}-4t^2\mathcal{L}\,,\qquad
\mathcal{R}=
\xi^3\mathcal{I} -6t\mathcal{V} -12t^2\mathcal{G} -8t^3\mathcal{L}\,,
&\nonumber
\end{eqnarray}
and by the second order supercharges $S_a$ 
and  the odd integrals $\mu_1=\frac{1}{2}\{\xi,\mathcal{P}_0\}\sigma_1
-i\frac{1}{2}[\xi,\mathcal{P}_0]\sigma_2-2tS_1$, $\mu_2=i\sigma_3\mu_1$,
$\kappa_1=\xi^2\sigma_1-4t\mu_1-4t^2S_1$, and  $\kappa_2=i\sigma_3\kappa_1$,
where $G_1^\alpha=D_1 G_0^\alpha D^\#_1$
is the Darboux-dressed free particle integral $G_0^\alpha$.
The resuting nonlinear (quadratic)  superalgebra is generated by ten even and ten odd integrals of the system 
$\mathcal{H}$ including  a trivial even central charge $\mathcal{I}$,
and represents  a nonlinearly super-extended Schr\"odinger algebra 
 with the $\mathfrak{osp}(2\vert 2)$ sub-superalgebra. The nontrivial bosonic generators
($\mathcal{L}$, 
$\mathcal{H}$, $\mathcal{G}$,
$\mathcal{P}_-$,
$\Sigma=\sigma_3$, $\mathcal{D}$,
$\mathcal{V}$,
$\mathcal{G}_-$,
$\mathcal{K}$,
$\mathcal{R}$)
are eigenstates of the dilatation generator $\mathcal{D}$,
$[\mathcal{D},\mathcal{O}]=is_{\mathcal{O}}\mathcal{O}$,
with the  eigenvalues given by 
$s_{\mathcal{O}}=(3/2,1,1/2,1/2,0,0,-1/2,-1/2,-1,-3/2)$.
Analogously, for  the fermionic generators
($\mathcal{S}_a$,
 $\mathcal{Q}_a$, $\mu_a$, $\lambda_a$,
 $\kappa_a$),  $s_{\mathcal{O}}=(1,1/2,0,-1/2,-1)$.
    The peculiarity of the nonlinear superalgebra, whose explicit form  is described in 
    \cite{JM2}, is 
    that the (anti)-commutators
of the generators of the $\mathfrak{osp}(2\vert 2)$  sub-superalgebra with any other generator is linear in
generators.  

 
 A simple example of the system in the phase of the partially broken 
 phase of the exotic nonlinear $\mathcal{N}=4$ 
 Poincar\'e supersymmetry is given by the Hamiltonian
 $ \mathcal{H}=\text{diag}\,(H^{\alpha_2}_1, H^{\alpha_1}_1)$
 composed from two Calogero systems regularized by  different 
 complex shifts   $\alpha_1>\alpha_2$.
 The  subsystems  $H^{\alpha_1}_1$ and
 $H^{\alpha_2}_1$
can be intertwined by the second
order differential operators 
 $D_{\alpha_1}D_{\alpha_2}^\#$
  and
 $D_{\alpha_2}D_{\alpha_1}^\#$
  via the `virtual'  free particle system, 
 $(D_{\alpha_1}D_{\alpha_2}^\#)H^{\alpha_2}_1=
H^{\alpha_1}_1(D_{\alpha_1}D_{\alpha_2}^\#)\,,$
$(D_{\alpha_2}D_{\alpha_1}^\#)H^{\alpha_1}_1=
H^{\alpha_2}_1(D_{\alpha_2}D_{\alpha_1}^\#).$
However, there also exists the first order intertwiners,
$D=\frac{d}{dx}+\mathcal{W}\,,$
$D^\#=-\frac{d}{dx}+\mathcal{W}\,,$
where
$\mathcal{W}=\frac{1}{\xi_1}-\frac{1}{\xi_2}-
\frac{1}{\xi_1-\xi_2},
$ $\xi_j=x+i\alpha_j$:
$DH^{\alpha_1}_1=H^{\alpha_2}_1D\,,$
$D^\#H^{\alpha_2}_1=H^{\alpha_1}_1D^\#.
$
They satisfy the relations
$D^\# D=H^{\alpha_1}_1-\Delta^2$,
$DD^\#=H^{\alpha_2}_1-\Delta^2$,
where 
$\Delta=(\alpha_1-\alpha_2)^{-1}$.
The supercharges  and Lax-Novikov integral of this extended 
system are
 \be\nonumber
Q_1= \left(
\begin{array}{cc}
0 & D   \\
D^\#  &  0
\end{array}
\right),\qquad
S_1= \left(
\begin{array}{cc}
0 & D_{\alpha_2}D^\#_{\alpha_1}   \\
D_{\alpha_1} D^\#_{\alpha_2}&  0
\end{array}
\right),\
\ee
 \be\nonumber
\mathcal{L}_1= \left(
\begin{array}{cc}
\mathcal{P}^{\alpha_2} & 0   \\
0&  \mathcal{P}^{\alpha_1}
\end{array}
\right),
\ee
$Q_2=\sigma_3Q_1$,
$S_2=\sigma_3s_1$.
They satisfy nontrivial superalgebraic relations
\be\nonumber
\{Q_a,Q_b\}=2\delta_{ab}(\mathcal{H}-\Delta^2)\,,\qquad
\{S_a,S_b\}=2\delta_{ab}\mathcal{H}^2\,,\qquad
\{Q_a,S_b\}=2\left(\epsilon_{ab}\mathcal{L}_1+i\delta_{ab}\Delta\mathcal{H}
\right)\,.
\ee
The exotic nonlinear supersymmetry here is in  the
 spontaneously partially broken phase:
 the doublet of the bound states 
 $\Psi^\pm_{0}=(D_{\alpha_2}1,\pm D_{\alpha_1}1)^t=
(-\xi^{-1}_2,\mp\xi^{-1}_1)^t$
of zero energy at the very edge of the fourfold degenerate continuous spectrum
are not annihilated by the first order supercharges,
 $Q_1\Psi^\pm_{0}=\pm i\Delta\Psi^\pm_0$.

In the case of the system
 $\mathcal{H}=\text{diag}(H^{\alpha_2}_1, H^{\alpha_1}_1)$,
its nonlinear superconformal algebra is more complicated 
\cite{JM2}.
The numbers of the even and odd generators are the same as in the previous example, 
but no odd fermionic generator
has a definite scaling dimension, i.e. is not an eigenstate of the dilatation operator
$\mathcal{D}$.  As a consequence,  the $\mathfrak{osp}(2\vert 2)$ superalgebra 
is not contained as a sub-superalgebra in this case.
 
It  is interesting to note that the two simplest $\mathcal{PT}$-regularized 
Calogero models $H^\alpha_\ell=-\frac{d^2}{dx^2}+\ell(\ell+1)\xi^{-2}$ with 
$\ell=1,2$  control stability properties of the 
$\mathcal{PT}$-regularized  kinks in the field-theoretical Liouville and $SU(3)$ 
conformal Toda systems
\cite{MatPly}.

\begin{figure}[htbp]
\begin{center}
\includegraphics[scale=0.5]{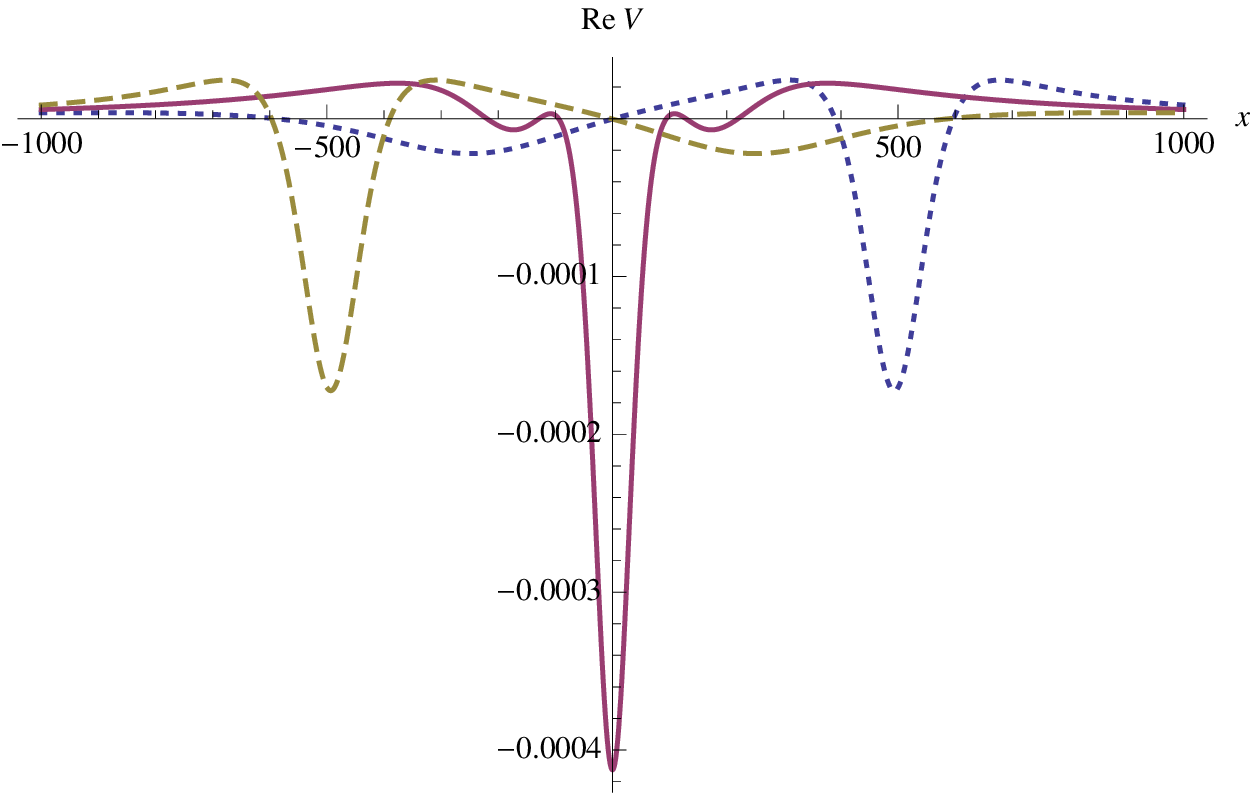}\includegraphics[scale=0.5]{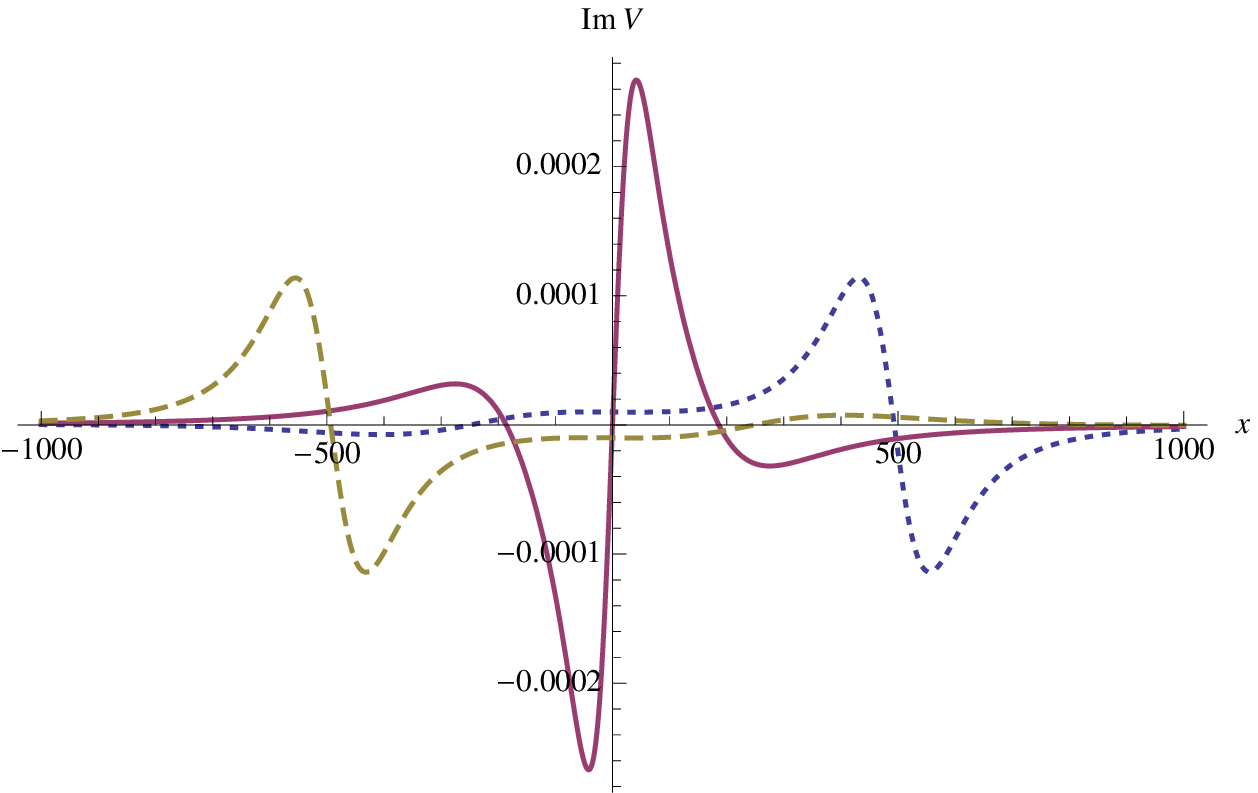}
\caption{  Evolution of real $v(x,\tau)$  (on the left) and imaginary $w(x,\tau)$ (on the right)  parts of
the potential  
$V_+(x;\alpha,\gamma(\tau,\nu))$
as a  complex $\mathcal{PT}$-symmetric solution 
of the KdV equation at  $\alpha=100$, 
 $\nu=5$;
dashed lines:  $\tau=-10^{7}$, continuous lines: 
$\tau=0$, dotted lines: 
$\tau=10^{7}$.
}\label{FigV1E}
\end{center}
\end{figure}

Consider now  the state $\psi_{\alpha,\gamma}^{(1)}={\gamma}{\xi^{-1}}+\xi^2$,
$\xi=x+i \alpha$,
$\alpha\in \R$, $\gamma=12\tau+i\nu\alpha^3$, $\nu\in(1,\infty)$, 
$\tau\in (-\infty,\infty)$, which 
is a linear combination of the bound state 
$\xi^{-1}$ of the system  $H^\alpha_1=-\frac{d^2}{dx^2}+\frac{2}{\xi^2}$ of zero eigenvalue and 
of its non-physical partner $\xi^2$ of the same zero energy.  Taking it as a seed 
state for the generalized Darboux transformation, 
we obtain a superpotential $\mathcal{W}^{(1)}_{\alpha,\gamma}=\frac{d}{dx}\left(\ln 
\psi_{\alpha,\gamma}^{(1)}\right)=-{\xi}^{-1}+3\xi^2(\xi^3+\gamma)^{-1}$,
and generate the super-partner systems $H_\pm=-\frac{d}{dx^2}+V_\pm$
given in a usual way by the potentials $V_\pm=(\mathcal{W}^{(1)}_{\alpha,\gamma})^2\pm
(\mathcal{W}^{(1)}_{\alpha,\gamma})'$ . This yields 
$V_+=2\xi^{-2}$, i.e. $H_+=H^\alpha_1$, and  $H_-=H^{\alpha,\gamma}_2=-\frac{d^2}{dx^2}+V_-$,
where
\begin{equation}\nonumber
 V_-=-2\left(\ln W(\xi,-\gamma+\xi^3)\right)''=
\frac{6}{\xi^2}-6\gamma\frac{4\xi^3+\gamma}{\xi^2(\xi^3+\gamma)^2}:=V(x;\alpha,\gamma(\tau,\nu))\,.
\end{equation}
The first equality with the Wronskian $W$means here that the system $H_-$ can also be produced
directly from the free particle system by taking as the set of the seed states 
for the generalized Darboux transformation the non-physical 
zero-energy eigenstate $\xi$ of the free particle and a linear 
combination $-\gamma+\xi^3$ 
of its zero-energy eigenstate $-\gamma$ and its Jordan state $\xi^3$,
$H_0\xi^3=-6\xi$.  As a function of $x$  and $\tau$, the potential
 $V(x;\alpha,\gamma(\tau,\nu)):=u(x,\tau)$ satisfies the complexified KdV equation
 $u_\tau-6uu_x+u_{xxx}=0$ being regular function for all values
 of  $x$ and $\tau$.  In the case $\alpha=0$, 
 potential $V$ takes the form of the well known singular rational  solution 
 $u(x,\tau)=6x\frac{x^3-24\tau}{(x^3+12\tau)^2}$ of the KdV equation.
 Note also that the potential $V(x;\alpha,\gamma(\tau,\nu))$ as a function 
 of $x$ satisfies simultaneously the higher stationary equation
 of the KdV hierarchy, $30u^2u_x-20u_xu_{xx}-10uu_{xxx}+u_{xxxxx}=0$.
 The real and imaginary parts of the potential $u(x,\tau)=v(x,\tau)+iw(x,\tau)$
 obey the system of the coupled nonlinear equations $v_\tau-3(v^2-w^2)_x+v_{xxx}=0$
and  $w_\tau-6(vw)_x+w_{xxx}=0$, and 
 represent some two-soliton waves.  For appropriately chosen parameters 
 $\alpha$ and $\nu$, they  reveal the behaviour  
 typical for extreme (rouge) waves 
 \cite{MatPly}, 
 see Figure \ref{FigV1E}.
   
    In conclusion we note that the rational extensions 
    of the harmonic oscillator, or of the 
    conformal Alfaro, Fubini, Furlan model (AFF) 
    \cite{deAFF,IvaKriLev1,SCM5}
    can be constructed  from the indicated systems
    by applying to them dual Darboux transformations
    with intertwining operators to be differential operators of the 
    even and odd orders
    \cite{CarPly1,CIPConf,InzPlyNu}. 
    The produced in such a way systems
    reveal a ``finite-gap" structure in their discrete spectra,
      but the dual Darboux schemes generate 
    from the harmonic oscillator or
    the AFF model  the pairs of the systems described by the Hamiltonians
    mutually shifted for a nonzero constant.
 As a consequence, instead of the Lax-Novikov type integrals, in this case 
    nontrivial ladder operators are generated, which allow to connect
    finite ``valence bands"  with equidistant infinite part of the spectrum.
    Using them, one can construct three pairs of  the ladder operators
    which encode the spectral peculiarities of  the  system and form a 
    complete spectrum-generating set of the ladder operators.
    Such rationally extended systems are characterized by 
    nonlinearly deformed extended conformal (Newton-Hooke) symmetry.
    They also can be related to the free particle  
    via the singular Darboux transformations and by the 
     conformal bridge construction described in a recent paper 
     \cite{ConBri}.
    Both Darboux and conformal bridge transformations 
      substantially use zero-energy eigenstates and 
    Jordan states corresponding to  zero-energy.
    
 
 Identifying a spatial reflection  $\mathcal{R}$ as a $\Z_2$-grading 
operator, 
the nonlinear $\mathcal{N}=2$ Poincar\'e  supersymmetry 
can be revealed  in many purely bosonic non-extended  quantum systems  
in the form of the bosonized supersymmetry 
\cite{PlyBoso,CorPly,Veron}.
In such systems, the Lax-Novikov  
integrals $\mathcal{P}$ play the role of the local supercharge, while the 
second supercharge $i\mathcal{R}\mathcal{P}$
is nonlocal due to the nonlocal nature of the reflection 
operator.  
Similarly, a hidden superconformal symmetry 
is identified in the quantum harmonic oscillator system 
\cite{InzPlyHid}.

 \vskip0.1cm
 
{\bf  Acknowledgements}


 The work was partially supported by the 
FONDECYT Project 1190842, the Project USA 1899,
and  by DICYT,  USACH.

\end{document}